\documentclass[preprint,showpacs,superscriptaddress,floatfix,prl,amssymb,amsmath]{revtex4}
\usepackage[dvips]{graphicx}
\usepackage{epsfig}
\usepackage{tabularx}
\usepackage{pifont}
\usepackage{color}
\usepackage[pdfpagemode=UseNone,colorlinks=true,linkcolor=blue,citecolor=blue]{hyperref}

\newcommand{\omegam}{\omega_{\mathrm{m}}}
\newcommand{\ud}{\mathrm{d}}
\newcommand{\barg}{\bar{g}}
\newcommand{\bxi}{\boldsymbol{\xi}}
\newcommand{\bphi}{\boldsymbol{\phi}}
\newcommand{\tX}{\widetilde{X}}
\newcommand{\tY}{\widetilde{Y}}
\newcommand{\tXmin}{\widetilde{X}_{\rm{min}}}
\newcommand{\Kbet}{K_{\beta}}
\newcommand{\Kth}{K_{\mathrm{th}}}
\newcommand{\kB}{k_{\mathrm{B}}}
\newcommand{\Smin}{S_{\mathrm{min}}}
\newcommand{\thmin}{\theta_{\mathrm{min}}}
\newcommand{\Smax}{S_{\mathrm{max}}}
\newcommand{\tlong}{\tau_{\mathrm{long}}}
\newcommand{\tshort}{\tau_{\mathrm{short}}}

\begin{document}

\title{Dynamical two-mode squeezing of thermal fluctuations in a cavity opto-mechanical system}

\author{A. Pontin}
\affiliation{Dipartimento di Fisica e Astronomia, Universit\`a di Firenze, Via Sansone 1, I-50019 Sesto Fiorentino (FI), Italy}
\affiliation{INFN, Sezione di Firenze}

\author{M. Bonaldi}
\affiliation{Institute of Materials for Electronics and Magnetism, Nanoscience-Trento-FBK Division,
 38123 Povo, Trento, Italy}
\affiliation{Istituto Nazionale di Fisica Nucleare (INFN), Trento Institute for Fundamental Physics and Application, I-38123 Povo, Trento, Italy}

\author{A. Borrielli}
\affiliation{Institute of Materials for Electronics and Magnetism, Nanoscience-Trento-FBK Division,
 38123 Povo, Trento, Italy}
\affiliation{Istituto Nazionale di Fisica Nucleare (INFN), Trento Institute for Fundamental Physics and Application, I-38123 Povo, Trento, Italy}

\author{L. Marconi}
\affiliation{Dipartimento di Fisica e Astronomia, Universit\`a di Firenze, Via Sansone 1, I-50019 Sesto Fiorentino (FI), Italy}

\author{F. Marino}
\affiliation{INFN, Sezione di Firenze}
\affiliation{CNR-INO, L.go Enrico Fermi 6, I-50125 Firenze, Italy}

\author{G. Pandraud}
\affiliation{Dept. of Microelectronics and Computer Engineering /ECTM/DIMES, Delft University of Technology, Feldmanweg 17, 2628 CT  Delft, The Netherlands}

\author{G. A. Prodi}
\affiliation{Istituto Nazionale di Fisica Nucleare (INFN), Trento Institute for Fundamental Physics and Application, I-38123 Povo, Trento, Italy}
\affiliation{Dipartimento di Fisica, Universit\`a di Trento, I-38123 Povo, Trento, Italy}

\author{P.M. Sarro}
\affiliation{Dept. of Microelectronics and Computer Engineering /ECTM/DIMES, Delft University of Technology, Feldmanweg 17, 2628 CT  Delft, The Netherlands}

\author{E. Serra}
\affiliation{Istituto Nazionale di Fisica Nucleare (INFN), Trento Institute for Fundamental Physics and Application, I-38123 Povo, Trento, Italy}
\affiliation{Dept. of Microelectronics and Computer Engineering /ECTM/DIMES, Delft University of Technology, Feldmanweg 17, 2628 CT  Delft, The Netherlands}

\author{F. Marin}
\email[Electronic mail: ]{marin@fi.infn.it}
\affiliation{Dipartimento di Fisica e Astronomia, Universit\`a di Firenze, Via Sansone 1, I-50019 Sesto Fiorentino (FI), Italy}
\affiliation{INFN, Sezione di Firenze}
\affiliation{CNR-INO, L.go Enrico Fermi 6, I-50125 Firenze, Italy}
\affiliation{European Laboratory for Non-Linear Spectroscopy (LENS), Via Carrara 1, I-50019 Sesto Fiorentino (FI), Italy}

\date{\today}
\begin{abstract}
We report the experimental observation of two-mode squeezing in the oscillation quadratures of a thermal micro-oscillator. This effect is obtained by parametric modulation of the optical spring in a cavity opto-mechanical system.  In addition to stationary variance measurements, we describe the dynamic behavior in the regime of pulsed parametric excitation, showing enhanced squeezing effect surpassing the stationary 3dB limit. While the present experiment is in the classical regime, our technique can be exploited to produce entangled, macroscopic quantum opto-mechanical modes.
\end{abstract}


\maketitle

Cavity opto-mechanics \cite{AspelRMP} is achieving several major experimental breakthroughs, including the optical cooling of opto-mechanical oscillators down to a thermal occupation number around or below unity \cite{oconnell10,teufel11,chan11,verhagen12,Purdy2012}. A great deal of attention  is now devoted to possible realizations of strongly non-classical states of macroscopic variables, that would open the way to new quantum information tools, as well as to crucial tests on the classical-to-quantum transition. In this framework, several techniques to obtain squeezing of the motion quadratures have been studied \cite{Clerk2008,Jahne2009,Mari2009,Bowen2011,Vanner2011,Asjad2014} and demonstrated on thermal oscillators \cite{Rugar1991,Briant2003,Hertzberg2009,Bowen2013,Vanner2013,Pontin_PRL2014,Poot2014}, including quantum non-demolition measurements with modulated \cite{Clerk2008,Hertzberg2009} and  pulsed  \cite{Vanner2011,Vanner2013} readout, and parametric excitation \cite{Bowen2011,Rugar1991,Briant2003,Bowen2013,Pontin_PRL2014,Poot2014}. Very recent reports describe the quantum squeezing of the motion of a cooled nano-oscillator exploiting the mechanical interaction with a microwave field \cite{Wollman2015,Pirkkalainen2015}. A different class of macroscopic, strongly non-classical systems can be obtained by involving more than one optical and/or mechanical mode. In particular, two mechanical modes have been exploited in opto-mechanical hybridization \cite{Shkarin} and are the basic ingredients of  two-mode mechanical squeezing, that is the subject of this work. 

A two-mode mechanical squeezed state is characterized by correlated fluctuations between one quadrature of the first oscillating mode and one of the second mode (as well as between the corresponding conjugate quadratures). In a quantum system, such correlations translate into entanglement between the two oscillators, i.e., the mechanical correspondent of optical "twin beams" generated in optical parametric amplifiers \cite{Gerry}. In a suitable combination of quadratures of the two oscillators, the fluctuations are reduced below their standard quantum level (or, in a classical system, below the thermal noise level). Such a reduction can be obtained by implementing a modulation of the spring constant at the sum of the resonance frequencies of the two oscillators.  Experimental demonstrations of fully mechanical two-mode squeezing of thermal fluctuations have been recently reported for nano-oscillators where the parametric modulation is obtained by electro-mechanic driving, either incorporating a piezoelectric transducer directly  into the mechanical elements of a device designed on purpose\cite{MahboobPRL2014}, or taking advantage from an accidental resonance with a mode of the substrate of a membrane oscillator \cite{Patil2014}. Here we study two-mode thermal squeezing in a cavity opto-mechanical system, exploiting parametric modulation of the optical spring. Our method, just based on the fundamental opto-mechanical interaction, does not require peculiar oscillator characteristics and can be extended to any cavity opto-mechanical system, including microwave driven nano-mechanical oscillators, and even in a moderately resolved sidebands regime. In addition to stationary variance and spectrally resolved measurements, we describe analyses of dynamic behavior in non-stationary conditions, showing enhanced squeezing effect surpassing the stationary limit determined by the onset of parametric instability. 

\begin{figure}[h]
\centering
\includegraphics[width=0.9\columnwidth]{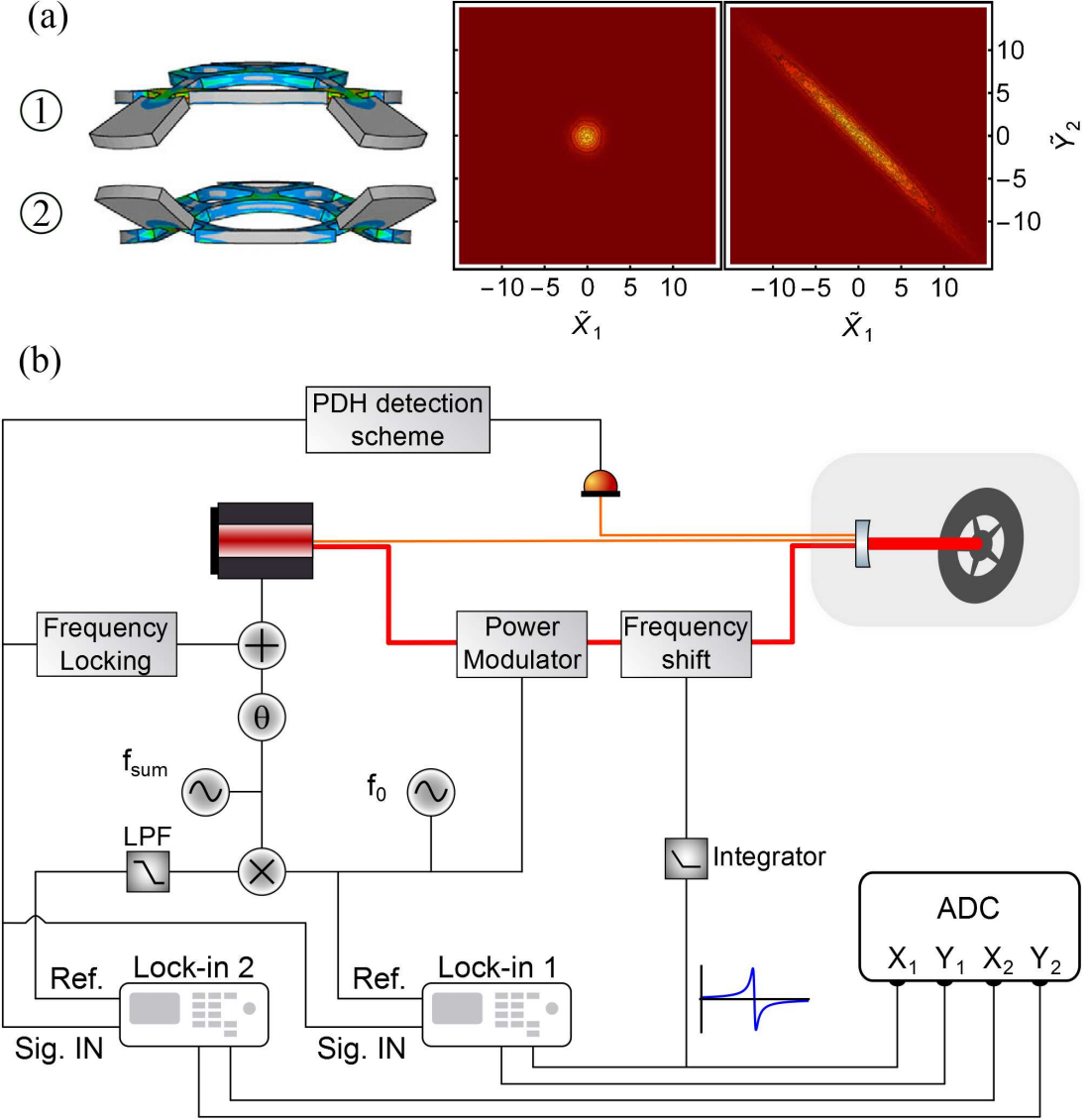}
\caption{(Color online) (a) Probability distributions in the ($\tY_1, \tX_2$) quadratures plane. Left: thermal distributions, without parametric excitation. Right: with parametric excitation, the fluctuations in the two quadratures are strongly correlated. $\tY_1$ and $\tX_2$ are normalized quadratures of two different oscillating modes, whose shape is shown by Finite Elements Model simulations on the left of the panels.  (b) Experimental setup. The first oscillation mode is parametrically locked at f$_0$ using the optical spring. The frequency f$_{\mathrm{sum}}$-f$_0$, corresponding to the resonance of the second mode, is selected after a mixer by the low-pass filter (LPF) and used as reference for the second lock-in amplifier.}
\label{fig_setup}
\end{figure}

Our silicon oscillator is made on the $70\,\mu$m thick device layer of a SOI wafer \cite{SerraJMM2013} and is composed of a central mass kept by structured beams \cite{SerraAPL2012}, balanced by four counterweights on the beams joints \cite{Borrielli2014}. On the surface of the central disk, a multilayer SiO$_2$/Ta$_2$O$_5$ dielectric coating forms a high reflectivity mirror. For the main oscillator mode, labeled \ding{172} in Fig. \ref{fig_setup}a (resonance frequency 172 kHz, effective mass $250 \mu$g), the motion of the mirror is balanced by the motion of the counterweights. As a result, the total recoil on the inner frame is null and the joints become nodal points. In a second oscillation mode, the counterweights move in the same direction of the mirror (mode \ding{173} in Fig. \ref{fig_setup}a), giving slightly different frequency  (225 kHz) and mass  ($100 \mu$g). Thanks to the overall design strategy, both modes reproducibly display a mechanical quality factor of about $Q \simeq 5 \times 10^4$ at room temperature, only limited by the thermoelastic dissipation in the substrate, and the balanced mode exhibits a $Q$ exceeding $10^6$ at cryogenic temperature \cite{PRApplied}. The oscillator reflecting surface is the end mirror of  a $L_{cav} = 1.27$~mm long, high Finesse ($\mathcal{F}=18000$, half-linewidth $\kappa/2\pi = 3.2$~MHz), over-coupled (input mirror transmission $\mathcal{T}=300$~ppm) Fabry-Perot cavity.  

In our setup  (Fig. \ref{fig_setup}b), two laser beams derived from the same Nd:YAG source are superimposed with orthogonal polarizations and coupled to the cavity. The first one is used  to obtain a signal proportional to the detuning, with an rf sidebands technique. This signal is used to lock the laser to the cavity resonance, with a bandwidth of about 10 kHz, and to measure the oscillator displacement $x(t)$. Furthermore, it is sent to two double-phase digital lock-in amplifiers whose outputs are acquired for the reconstruction of the motion of the oscillating modes. The second, more powerful beam, with a controllable frequency shift with respect to the previous one, sets and controls the optical spring. 

The intracavity radiation pressure force, depending on the detuning and therefore on the oscillator position, originates a so-called \emph{optical spring} \cite{AspelRMP} that modifies the mechanical resonance frequency. Moreover, due to the finite cavity field buildup time, it changes the effective damping of the oscillator motion, resulting in modal optical cooling (cold damping) when the input radiation frequency $\omega_L/2\pi$ is red shifted with respect to a cavity resonance $\omega_c/2\pi$. In the \emph{bad cavity} limit and for small detuning (i.e., when both the opto-mechanical frequencies and $\Delta = \omega_L-\omega_c$ are well below $\kappa$), 
the optical spring constant can be approximated as $K_{\mathrm{opt}} \simeq \frac{4 \omega_L P}{L_{cav} \kappa^2 c}\,\Delta$, where $P$ is the intracavity power. In the same conditions, the effective linewidth becomes $\,\,\gamma_{\mathrm{eff}}=\gamma_m+\gamma_{\mathrm{opt}}\,\,$ with
$\gamma_{\mathrm{opt}} \simeq -\frac{2 K_{\mathrm{opt}}}{m \kappa}$ ($m$ is the mass, $\omegam/2 \pi$ the resonance frequency and $\gamma$ the linewidth of the free oscillator) \cite{AspelRMP,Arcizet2006}. We exploit the optical spring to stabilize the frequency of the first oscillation mode, by slightly exciting it with an amplitude modulation on the strong laser beam, comparing its motion with the reference oscillator on the first lock-in, and phase-locking its oscillation to the reference by acting on the optical spring through the pump laser detuning. Such parametric locking \cite{Pontin_PRL2014} is also indirectly efficient on the second mode, since it stabilizes the whole optomechanical system.   

We now consider two oscillation modes, both changing the cavity length, with effective displacement $x_i(t)$ ($i = 1,2$), momentum $p_i(t)$, resonance frequency $\omega_i$ and mass $m_i$. It is convenient to observe the oscillators in their respective rotating frames. We thus analyze the behavior of the slowly varying quadratures, defined as $X_i = x_i \cos \omega_i t - \frac{p_i}{m_i\omega_i} \sin \omega_i t$ and $Y_i = x_i \sin \omega_i t + \frac{p_i}{m_i\omega_i} \cos \omega_i t$. We remark that the parametric locking above described acts on one quadrature of the first oscillator (namely, $X_1$), while the orthogonal one remains unaffected. Therefore, in the following we just compare the model with the experimental data concerning the $Y_1$ quadrature, even if for the theory (neglecting active parametric locking) the two quadratures are formally equivalent.  

In our experiment, the laser detuning is modulated at $\omega_1+\omega_2$ such that $\Delta \to \Delta [1+ \beta\,\cos (\omega_1+\omega_2)t]$ (with $\beta \ll 1$). As a consequence, the optical spring constant is also modulated, and the resulting force is $-(x_1+x_2)\,\Kbet\,\cos (\omega_1+\omega_2)t$, with $\Kbet = K_{\mathrm{opt}} \beta$. 

Writing the modulated force in terms of the quadratures, we derive that its quasi-resonant terms are associated to the two-mode squeezing Hamiltonian $H_I = \frac{\Kbet}{4} \left(X_1X_2 - Y_1Y_2 \right)$, that couples the quadrature $X_1$ with $Y_2$ and $Y_1$ with $X_2$. As a consequence, the two-mode system can be decomposed into two equivalent independent subsystems. The Langevin equations governing their motion can be written as
\begin{equation}
\dot{\bf{X}} + \bf{A} \bf{X} = \boldsymbol{\xi}
\label{sistema}
\end{equation}
where ${\bf X}=(X,Y)$ can be either ($X_1, Y_2$) or ($Y_1, X_2$),  $\boldsymbol{\xi} = (\sqrt{S_{\xi_1}} f_1, \sqrt{S_{\xi_2}} f_2 )$ with $\langle f_i(t) f_j (t')\rangle = \delta_{ij}\delta(t-t') $, $S_{\xi_i} = \sqrt{\frac{S_{Fi}}{2}} \frac{1}{m_i\omega_i}$ and $S_{Fi}$ is the force spectral density acting on the $i$-th oscillation mode. The system matrix is
\begin{equation}
\bf{A}=\left(\begin{array}{cc}
\frac{\gamma_1}{2} & g_1 \\
 g_2  & \frac{\gamma_2}{2}
 \end{array} \right)
 \label{eqA}
 \end{equation}
with $g_i = \Kbet/4\omega_i m_i$.

Stationary, null-average solutions are stable if $\barg=\Kbet/\Kth<1$ where the threshold $\Kth$, calculated by nulling the determinant of ${\bf A}$, is $\Kth = 2\sqrt{\gamma_1 \gamma_2 m_1 m_2 \omega_1 \omega_2}$. In this situation, it is interesting to evaluate the peak spectral densities of the quadratures (at $\omega = 0$). Indeed, they are proportional to the variance measured after integration over periods much longer than $1/\gamma$, that is a frequent experimental procedure.

In the simplified case of oscillators with identical susceptibilities (with mass $m$, resonance frequency $\omega_0$, damping rate $\gamma$, force spectral density $S_{F}$), the eigenvalues of ${\bf A}$ are simplified to $\lambda_{\pm} = \frac{\gamma}{2} (1 \pm \barg)$ and the respective eingenvectors are $X_{\pm} = (X \pm Y)/\sqrt{2}$. Below threshold, the spectra of the fluctuations around the stationary solutions are 
\begin{equation}
S_{X_\pm}(\omega)=\frac{2\lambda_\pm}{\omega^2+\lambda_\pm^2}\,\sigma_\pm^2
\end{equation}
where the variances are $\sigma^2_\pm = (\sigma^0)^2\,\gamma/2 \lambda_\pm$ with $(\sigma^0)^2 = S_F/(2 \gamma m^2\omega_0^2)$. On the combined quadrature $X_+$,  the peak spectral density $S_{X_+}(0)$ is reduced by a factor of $(1+\barg)^2$. The total variance is accordingly reduced by a factor of $(1+\barg)$, with a maximum reduction of respectively -6dB and -3dB when approaching the threshold. The quadrature $X_-$ has maximal fluctuations, with peak spectral density and variance increased respectively by $1/(1-\barg)^2$ and $1/(1-\barg)$ with 
respect to their values in the absence of parametric modulation. 

In the realistic case of different parameters between the two oscillation modes, to maintain meaningful combined quadratures it is useful to first normalize each quadrature to its standard value, calculated without parametric excitation. Therefore, we calculate the normalized quadratures $\tX_i \propto X_i$ and $\tY_i \propto Y_i$ in order to have either $S^0_{\tX_i}(0) = S^0_{\tY_i}(0) = 1$ or $(\sigma^0_{\tX_i})^2 = (\sigma^0_{\tY_i})^2 = 1$, according to the chosen indicator (we use the "0" superscript for the quantities measured without parametric excitation, i.e., at $g = 0$).  Eqs. (\ref{sistema}-\ref{eqA}) remain formally the same, with a different physical meaning of the parameters $g_i$ and $S_{\xi_i}$ \cite{suppl}. 
The general combined quadrature is now defined as $X_\theta = \tX \cos \theta + \tY \sin \theta$, with spectral density $S_\theta$ and total variance $\sigma^2_\theta$. Without parametric excitation, we have symmetric distributions with either $S^0_\theta (0)=1$, or $(\sigma^0_\theta)^2=1$, for any value of $\theta$, as shown in the left panel of Fig. \ref{fig_setup}a. 
With parametric excitation, the two quadratures corresponding to eigenvectors of ${\bf A}$ still maintain a Lorentzian spectrum, with width $\lambda_\pm$. However, differently from the case of identical oscillators, they do not correspond to extrema in the quadrature fluctuations. Either $S_\theta (0)$ or $\sigma^2_\theta$ should be minimized with respect to $\theta$ to derive the combined quadrature exhibiting the largest squeezing, in correspondence of $\thmin$. The probability distribution with parametric driving is shown in the right panel of Fig. \ref{fig_setup}a. The strong anti-correlated fluctuations in the two quadratures are clearly revealed by the elongated shape of the graph. In a narrow interval of $\theta$, close to $\pi/4$, the fluctuations of $X_{\theta}$ are below the thermal ones. 

It is important to remark that the described picture assumes an exact detection phase. In the realistic case of (small) fluctuations $\Delta\theta$ during the measurement time, the measured minimal spectral density is $\,\,\Smin+\langle (\Delta\theta)^2 \rangle \Smax$ where $\Smin = S_{\thmin}(0)$ and $\Smax = S_{\thmin+\pi/2}(0)$ (a similar expression holds for the variance). This feature is crucial to describe the experimental results. The expected minimal spectral density now reaches a minimal value before the threshold gain, followed by a deterioration of the squeezing. 

\begin{figure}[h]
\centering
\includegraphics[width=0.9\columnwidth]{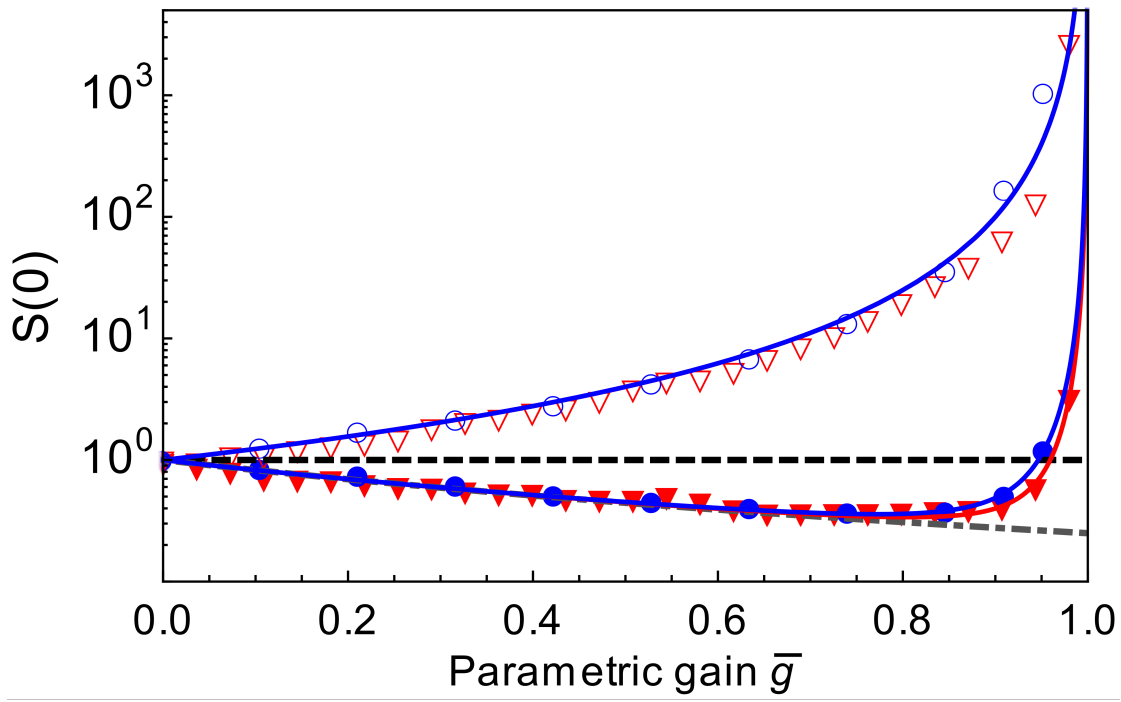}
\caption{(Color online) Experimental measurements of the spectral densities $\Smin$ (full symbols) and  $\Smax$ (empty symbols) as a function of the parametric gain, together with the corresponding theoretical predictions (solid lines). The signal from the four quadratures are preliminarily normalized in order to have unity spectral density in the absence of parametric excitation. Different colours (grey tones) corresponds to two data sets, taken with different levels of optical cooling. A dash-dotted line shows the prediction for a perfectly stable and optimized detection phase.}
\label{fig_stazionari}
\end{figure}
In Fig. \ref{fig_stazionari} we show our experimental measurement of $\Smin$ and  $\Smax$ as a function of the parametric gain, together with the corresponding theoretical predictions. The experimental values of $\Smin$ and $\thmin$ are obtained by globally minimizing the peak spectral density of the combined quadrature $\,\,(\tX_1 \cos \phi_1+\tY_1 \sin \phi_1) \cos\theta+(\tX_2 \sin \phi_2+\tY_2 \cos \phi_2) \sin\theta\,\,$ with respect to $\phi_1$, $\phi_2$ and $\theta$, to account for possible errors in the fixed lock-in reference phases.

As expected and already observed \cite{MahboobPRL2014,Patil2014}, the squeezing does not surpass -6dB before the onset of parametric instability. Moreover, $\Smin$ reaches a minimum before the threshold gain, in contrast with the expected behaviour for a perfect and stable detection phase, displayed in the picture with a grey dash-dotted line. This feature, already observed in recent experiments \cite{MahboobPRL2014}, is well reproduced by the complete model with the parameters of our different oscillating modes. The theoretical values are calculated using independently measured oscillators parameters, and just fitting to the experimental data the threshold parametric gain, whose independent estimation is not enough accurate, and the phase variance $ \langle (\Delta\theta)^2 \rangle $ that results to be around $10^{-3}$ rad$^2$. 

Above threshold, the system admits no stable solutions and therefore no stationary spectra. However, one can still calculate the behavior of the variance in measurements of a quadrature, performed at the time $t$ after the switch-on of the parametric modulation \cite{suppl}. 

To start describing the physics involved, we present again the case of identical oscillators. As before, the extrema of the variance are found for $\theta=\pm\pi/4$ at any time $t$, i.e., in correspondence of the eigenvectors of $\bf{A}$, with $\sigma^2_{\mathrm{min}} \equiv \sigma^2_+$ and $\sigma^2_{\mathrm{max}} \equiv \sigma^2_-$. The respective temporal evolutions are $ \,\,\sigma^2_\pm(t) = (\sigma^0)^2 [\exp(-2\lambda_\pm t)+\frac{\gamma}{2\lambda_\pm}(1-\exp(-2\lambda_\pm t))]$. For $\barg>1$, $\sigma^2_{\mathrm{min}}$ approaches $1/(1+\barg)$ (i.e., a squeezing exceeding -3dB) with a time constant $\tshort=1/2\lambda_+ = 1/\gamma(1+\barg)$, while $\sigma^2_{\mathrm{max}}$ exponentially diverges with the longer time constant $\tlong=1/2|\lambda_-| = (\sigma^0)^2/\gamma(\barg-1)$, exhibiting the critical slowing down at threshold. If the system dynamic range is large enough, it can be left evolving after the switch-on of the parametric excitation for few $\tshort$, reaching therefore a stronger two-mode squeezing with respect to the stationary situation. 
  
For different oscillators, the variances $\sigma_{\pm}$ in the combined quadratures corresponding to eigenvectors of $\bf{A}$ still evolve exponentially with time constants $1/2\lambda_\pm$, while for any other quadrature we expect the combination of exponentials with time constants $1/2\lambda_+$, $1/2\lambda_-$, and $1/(\lambda_++\lambda_-)$. As in the stationary case, minimal fluctuations are no more found for the eigenvector corresponding to $\lambda_+$ (except for asymptotically long $t$, as we will see). Moreover, here the optimal phase $\theta$ depends on $t$, as well as on $\barg$. 

\begin{figure*}[h]
\centering
\includegraphics[width=0.98\columnwidth]{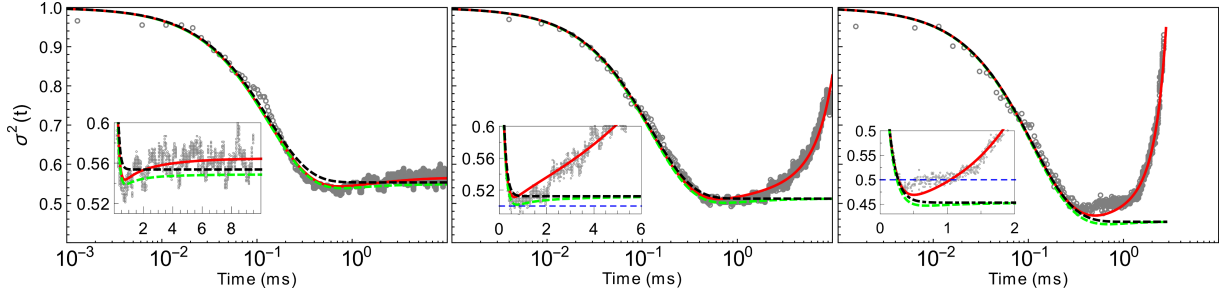}
\caption{(Color online) Time evolution of the experimental minimal variance $\sigma^2_{min}$ after the onset of the parametric modulation (symbols), together with the prediction of the model (solid lines), for different modulation amplitudes. The normalized parametric gains $\bar{g}$ extracted from the fits are $\bar{g} = 0.92$ (a), $\bar{g} = 1.06$ (b), $\bar{g} = 1.30$ (c). The second fit parameter is the phase noise, that results to be $\Delta \theta_{\mathrm{rms}} = 0.041$ rad (a), $\Delta \theta_{\mathrm{rms}} = 0.045$ rad (b), $\Delta \theta_{\mathrm{rms}} = 0.070$ rad (c). The variance is measured over 5000 consecutive bursts, repeated at the rate of 50 Hz. The duty cycle is 50$\%$ for panels (a) and (b), and 15$\%$ for panel (c). Light (green) dashed lines: predicted evolution in the absence of phase noise ($\Delta \theta = 0$). Black dashed lines: predicted evolution of $\sigma^2_+$.}
\label{fig_bursts}
\end{figure*}
We have applied to our system periodic bursts of parametric modulation. The bursts are long enough to observe the system evolution, and are repeated at the rate of 50 Hz. The four output signals of the two lock-in amplifiers (integrated with a time constant of 40 $\mu$s) are sampled at $1/\Delta t = 200$ kS/s, then sectioned synchronously with the periodic bursts, so that each section starts at the beginning of a burst. We note that the sampling rate is much larger than $\lambda_{\pm}$. For each $n$-th data sample, we calculate variances and cross-correlations between the four signals, at the time $n \Delta t$ after the onset of the parametric modulation, over 5000 consecutive bursts. The minimal variance  $\sigma^2_{\mathrm{min}}$ is obtained by global minimization with respect to the three phases $\phi_1$, $\phi_2$ and $\theta$. In Fig. \ref{fig_bursts} we report the time evolution of $\sigma^2_{\mathrm{min}}$, together with the prediction of the model, for three levels of gain $\bar{g}$ below, near and above threshold. As expected, in the last case the variance falls below the stationary limit of $\sim 0.5$. The effect of the phase fluctuations $\Delta\theta$ is even more crucial than in stationary conditions. The measurable minimal variance presents a minimum for an optimal value of $t$, after which it starts increasing due to the mixing with the exploding quadrature. We also remark that, in the absence of phase fluctuations, there is a qualitative difference between the two regimes $\bar{g}<1$ and $\bar{g}>1$. In the former case, the minimal variance is always lower than $\sigma^2_+$, as in the stationary case (see Fig. \ref{fig_bursts}). Above threshold, instead, it asymptotically converges to $\sigma^2_+$, and the combined quadrature with minimal fluctuations tends to coincide with the stable eigenvector.   

In conclusion, we have prepared an optomechanical system in a two-mode squeezed thermal state. Some features recently observed in purely mechanical and electro-mechanical systems, such as the deterioration of the squeezing before the parametric threshold, are accurately observed and well reproduced by a model considering different mode parameters. Moreover, we analyze the dynamical evolution after the onset of the parametric modulation and show that the -3dB limit in the squeezing of the two-mode variance, rigorous in stationary conditions for identical oscillators, is overcome for a bit of time. Even in this case, our model well reproduces the experimental data, and underlines the crucial role of phase fluctuations. Indeed, both the achievable minimal uncertainty and its temporal stability are actually determined by such phase noise.

Our system operates with low optical and mechanical losses and implements optical cooling and frequency control, with a setup similar to those employed to prepare the oscillator close to its ground state. In such conditions, our technique could produce entangled states of the mechanical oscillation modes \cite{Johansson2014,Li2015}, an intriguing result in the field of quantum macroscopic systems. The model here presented neglects the field fluctuations, privileging a clearer focus on the main features that are demonstrated experimentally. However, a complete quantum analysis shows that peculiar quantum effects can be created involving the electromagnetic  field, efficiently coupled to the cavity and detected. Thanks to its flexibility, our method opens the way to more elaborated schemes including more than two oscillating modes, overcoming the experimentally challenging requirement of equal oscillation frequencies, and leading to experimental studies of complex multipartite entanglement in macroscopic systems. The experimental techniques here described can also be applied to different devices embedded in the same optical cavity, or even in separated cavities sharing the same optical field \cite{Jockel}, thus producing well separated, entangled macroscopic oscillators, particularly interesting for investigating quantum decoherence.

This work has been supported by MIUR (PRIN 2010-2011 and QUANTOM project) and by INFN (HUMOR project). A.B. acknowledges support from the MIUR under the "FIRB-Futuro in ricerca 2013" funding program, project code RBFR13QUVI.

\section{Supplemental Material - Dynamical two-mode squeezing of thermal fluctuations in a cavity opto-mechanical system}

\subsection{The system of two different oscillators}

The quantum Langevin equations for the position $x(t)$ and the momentum $p(t)$ of a mechanical oscillator with mass $m$, damping rate $\gamma$ and resonance frequency $\omega_0$ are  
\begin{eqnarray}
\dot{x} - p/m = 0\\
\dot{p} + m\omega_0^2 \,x + m\gamma \,\dot{x} = \sqrt{S_F} f(t)
\end{eqnarray}
with $\langle f (t) f (t')\rangle = \delta(t-t') $ (valid in the limit of high mechanical quality factors $Q$ \cite{Giovannetti}). The force spectral densities can be written as $S_F \simeq \frac{\hbar m \omega_0^2}{Q}(2\bar{n}+1) $ where $\bar{n} = [\exp (\hbar\omega_0/\kB T)-1]^{-1} $ is the thermal mean occupation number. In the limit of high temperature ($\kB T \gg \hbar \omega_0$), valid for the present experiment, we obtain the classical thermal noise spectral density $S_F \simeq 2 \kB T m\omega_0/Q$, while in the opposite case of low occupation number ($\bar{n} \ll 1$) the quantum noise is $S_F \simeq 2 \hbar m \omega_0^2/Q$. If the oscillator is embedded in an optical cavity, both its resonance frequency and its damping rate are modified by the opto-mechanical interaction, therefore $\omega_0$ and $\gamma$ must be intended as effective parameters of the opto-mechanical system. The most meaningful effect is on the damping rate (that is usually strongly modified), therefore we keep distinguished the oscillator effective resonance linewidth $\gamma$ from its natural linewidth $\omega_0/Q$. In the classical limit, valid in our experiment, the additional noise source due to the quantum fluctuations of the radiation can be neglected, and the oscillator can be considered at thermal equilibrium with an effective temperature $T_{eff} = T \omega_0/\gamma Q$. When the quantum regime is approached by optical cooling ($T_{eff}$ approaching $\hbar \omega_0/\kB$), field fluctuations become important and a limit effective temperature is achieved. Our simplified calculations remain meaningful by replacing $\frac{\omega_0}{Q} \to \gamma$ and $T \to T_{eff}$.

It is useful to describe the motion of the oscillator using the slowly-varying quadratures $X$ and $Y$, defined as $X = x \cos \omega_0 t - \frac{p}{m\omega_0} \sin \omega_0 t$ and $Y = x \sin \omega_0 t + \frac{p}{m\omega_0} \cos \omega_0 t$. Their evolution equations (neglecting the terms rotating at $\sim 2 \omega_0 t$) are
\begin{eqnarray}
\dot{X} = -\frac{\gamma}{2} X + \sqrt{\frac{S_F}{2}}\frac{1}{m \omega_0} \,f^{(1)} \\
\dot{Y} = -\frac{\gamma}{2} Y + \sqrt{\frac{S_F}{2}}\frac{1}{m \omega_0} \,f^{(2)}
\end{eqnarray}
with $\langle f^{(i)} (t) f^{(j)} (t')\rangle = \delta_{ij}\delta(t-t') $.

We now consider the system of two different oscillators (singled out by the subscripts ''1'' and ''2''), coupled by a force of the form 
$ -(x_1+x_2)\,\Kbet\,\cos (\omega_1+\omega_2)t $.
Among the different force terms that are obtained by writing $x_1$ and $x_2$ in terms of the respective quadratures and expanding the periodic functions, the relevant (quasi-resonant) ones are $\frac{\Kbet}{2} \, (Y_2 \sin \omega_1 t -X_2 \cos \omega_1 t )$ for the first oscillator and $\frac{\Kbet}{2} \, (Y_1 \sin \omega_2 t -X_1 \cos \omega_2 t )$ for the second one. The evolution equations for the quadratures, neglecting again the force terms oscillating at twice the resonance frequency, become
\begin{eqnarray*}
m_1\omega_1 \left(\partial_t + \frac{\gamma_1}{2}\right) X_1 +\frac{\Kbet}{4} Y_2 = \sqrt{\frac{S_{F1}}{2}} f^{(1)}_{1} \\
m_1\omega_1 \left(\partial_t + \frac{\gamma_1}{2}\right) Y_1 +\frac{\Kbet}{4} X_2 = \sqrt{\frac{S_{F1}}{2}} f^{(2)}_{1} \\
m_2\omega_2 \left(\partial_t + \frac{\gamma_2}{2}\right) X_2 +\frac{\Kbet}{4} Y_1 = \sqrt{\frac{S_{F2}}{2}} f^{(1)}_{2} \\
m_2\omega_2 \left(\partial_t + \frac{\gamma_2}{2}\right) Y_2 +\frac{\Kbet}{4} X_1 = \sqrt{\frac{S_{F2}}{2}} f^{(2)}_{2}
\end{eqnarray*}
where the stochastic force sources have correlation functions $\langle f_i^{(a)}(t) f_j^{(b)} (t')\rangle = \delta_{ij}\delta_{ab}\delta(t-t') $. 

The coupling terms are associated to the Hamiltonian $H_I = \frac{\Kbet}{4} \left(X_1X_2 - Y_1Y_2 \right)$. Introducing the ladder operators in the rotating frame as $\hat{a}_i = 
\sqrt{\frac{m_i \omega_i}{2 \hbar}}\left(X_i + i Y_i \right)$, we can also write the two-mode squeezing Hamiltonian in the form $H_I = \hbar \barg \frac{\sqrt{\gamma_1 \gamma_2}}{2}\left(\hat{a}_1\hat{a}_2+\hat{a}^{\dagger}_1\hat{a}^{\dagger}_2\right)$ \cite{Gerry} where $\barg = \Kbet/2\sqrt{\gamma_1 \gamma_2 m_1 m_2 \omega_1 \omega_2}$ is the parametric gain normalized to threshold. $H_I$ couples the quadrature $X_1$ with $Y_2$ and $Y_1$ with $X_2$. As a consequence, the two-mode system can be decomposed into two equivalent independent sub-systems. The Langevin equations governing the motion of each sub-system can be written as

\begin{equation}
\dot{\bf{X}} + \bf{A} \bf{X} = \boldsymbol{\xi}
\label{sistema}
\end{equation}
where ${\bf X}=(X,Y)$ can be either ($X_1, Y_2$) or ($Y_1, X_2$),
\begin{equation}
\boldsymbol{\xi} = \left( \begin{array}{c}
\sqrt{\frac{S_{F1}}{2}} \frac{1}{m_1\omega_1} f_1 \\ 
\sqrt{\frac{S_{F2}}{2}} \frac{1}{m_2\omega_2} f_2 
\end{array} \right) 
\end{equation}
with $\langle f_i (t) f_j (t')\rangle = \delta_{ij}\delta(t-t') $ and the system matrix is
\begin{equation}
\bf{A}=\left(\begin{array}{cc}
\frac{\gamma_1}{2}\, & \, g_1 \\
 g_2 \, &\, \frac{\gamma_2}{2}
 \end{array} \right)
 \label{eqA}
 \end{equation}
with $g_i = \frac{\Kbet}{4\omega_i m_i}$. The eigenvalues of  ${\bf A}$ are
\begin{equation}
\lambda_\pm = \frac{1}{2}\left[\frac{\gamma_1+\gamma_2}{2} \pm \sqrt{\left( \frac{\gamma_1-\gamma_2}{2}\right)^2 + \gamma_1 \gamma_2 \,\barg^2}\right]  
\label{eigen}
\end{equation}
and they are simplified to $\lambda_{\pm} = \frac{\gamma}{2} (1 \pm \barg)$ for identical oscillators.

\subsection{Stationary solutions}

\begin{figure}[h]
\centering
\includegraphics[width=0.9\columnwidth]{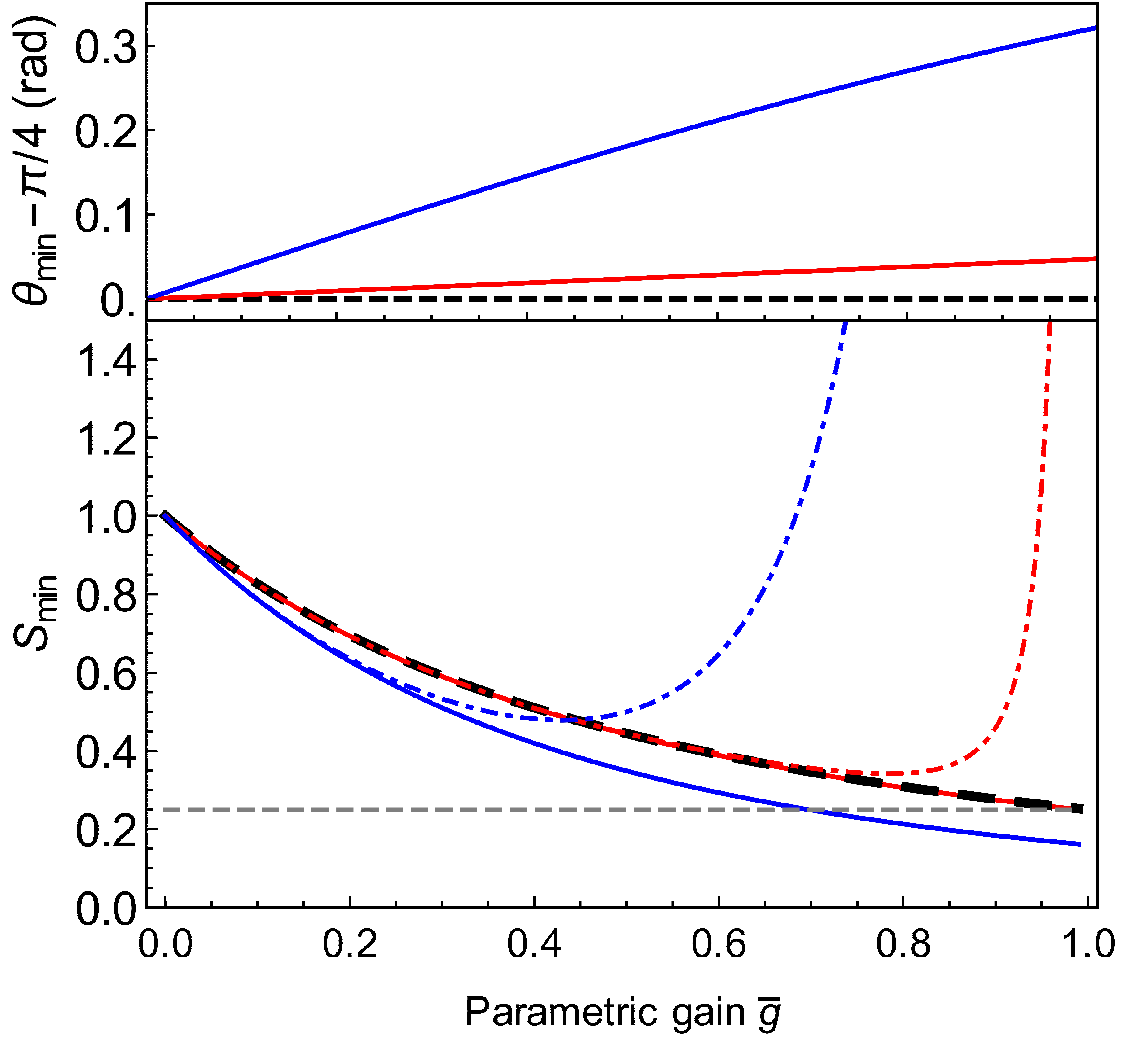}
\caption{Lower panel: with solid lines, we show the minimal spectral density $\Smin$ that can be obtained for the asymmetry parameter $\alpha = 1.1$ (red), corresponding to our experimental parameters, and with $\alpha = 2$ (blue). The black dashed line shows $\Smin$ when $\alpha = 1$ (i.e., for identical oscillators). With dashed-dotted lines, we report the fluctuations in the antisymmetric quadrature (with $\theta = \pi/4$). Upper panel: optimal phase $\thmin$ when $\alpha = 1.1$ (red) and $\alpha = 2$ (blue).}
\label{fig1}
\end{figure}

Below threshold, we can calculate stationary spectral densities and variances for any linear combination of the quadratures $X$ and $Y$. We start by calculating the spectral densities at vanishing frequency. At this purpose, we neglect the first term on the left of Eq. (\ref{sistema}). We normalize the quadratures in order to have unitary spectral density in the absence of parametric excitation (i.e., for $g_i = 0$). We define therefore $\tX = X/\sqrt{S^0_X}$ and $\tY = Y/\sqrt{S^0_Y}$  with $S^0_X = 2 S_{F1}/(m_1 \omega_1 \gamma_1)^2$ and 
$S^0_Y = 2 S_{F2}/(m_2 \omega_2 \gamma_2)^2$. Eq. (\ref{sistema}) can be written in the simple form
\begin{equation}
\left(\begin{array}{cc}
1 & \barg \alpha \\
 \barg/\alpha  & 1
 \end{array} \right)
\left( \begin{array}{c}
\tX \\ 
\tY 
\end{array} \right) 
= \left( \begin{array}{c}
f_1 \\ 
f_2 
\end{array} \right)
\label{sistema2}
\end{equation}
where we have defined a dimensionless asymmetry parameter $\alpha = \sqrt{\frac{S_{F2} m_1 \omega_1 \gamma_1}{S_{F1} m_2 \omega_2 \gamma_2}}$. In the classical (i.e., high temperature) limit, the asymmetry parameter can be written as $\sqrt{\gamma_1 Q_1/\gamma_2 Q_2}$, while in the strongly quantum regime (with $\frac{\omega_i}{Q_i} \to \gamma_i$ and $\bar{n} \ll 1$), $\alpha$ tends to 1 even for different oscillators. The solution of Eq. \ref{sistema2} is
\begin{equation}
\begin{array}{rcl}
\tX &=& \frac{f_1-\barg \alpha \, f_2}{1-\barg^2} \\
\tY &=& \frac{f_2-\barg/\alpha \, f_1}{1-\barg^2} 
\end{array}
\end{equation}
and has power spectral densities
\begin{equation}
\begin{array}{rcl}
S_{\tX \tX} &=& \frac{1+\barg^2 \alpha^2}{(1-\barg^2)^2}  \\
S_{\tY \tY} &=& \frac{1+\barg^2/\alpha^2}{(1-\barg^2)^2}  \\
S_{\tX \tY} = S_{\tY \tX} &=& -\frac{\barg}{(1-\barg^2)^2}\left( \alpha + \frac{1}{\alpha} \right)
\end{array}
\label{St}
\end{equation}
The minimal spectral density $\Smin$ corresponds to the quadrature $\tXmin = \tX \cos (\thmin) + \tY \sin (\thmin)$, with 
\begin{equation}
\thmin = \frac{1}{2} \arctan \frac{S_{\tX \tY}+S_{\tY \tX}}{S_{\tX \tX}-S_{\tY \tY}}
\label{thmin0}
\end{equation}  
and is equal to 
\begin{equation}
\Smin = \frac{1}{2} \left( S_{\tX \tX}+S_{\tY \tY}-\sqrt{(S_{\tX \tX}-S_{\tY \tY})^2+(S_{\tX \tY}+S_{\tY \tX})^2} \right)  \, .
\label{Smin0}
\end{equation}
Using Eqs. (\ref{St}), we obtain the explicit expressions for $\thmin$ and $\Smin$: 
\begin{equation}
\thmin = \frac{\pi}{2}-\frac{1}{2} \arctan \frac{2 \alpha}{\barg (\alpha^2-1)}
\label{thmin}
\end{equation}  
and  
\begin{equation}
\Smin = \frac{1}{(1-\barg^2)^2} \left( 1+\frac{\barg^2}{2}\left(\alpha^2+\frac{1}{\alpha^2}\right)-\sqrt{\frac{\barg^4}{4}\left( \alpha^2 - \frac{1}{\alpha^2} \right)^2 + \barg^2 \left( \alpha + \frac{1}{\alpha} \right)^2} \right)   \, .
\label{Smin}
\end{equation}
With our normalization, $\Smin$ is below its value with equal oscillators (i.e., $\Smin < 1/(1+\barg)^2$) whenever $\alpha \ne 1$, as shown in Fig. \ref{fig1}. We also underline in the same figure that measuring the fluctuations in the antisymmetric quadrature (i.e., with $\theta = \pi/4$) is not the optimal choice whenever $\alpha \ne 1$. In our experiment, $\alpha \lesssim 1.1$ and the difference between $\Smin$ and $1/(1+\barg)^2$ is small (e.g., 0.248 with respect to 0.25 for $\barg \to 1$), well below the experimental resolution. However, the accurate choice of $\thmin$ remains a crucial requirement even for such a weak asymmetry as soon as $\barg > \sim 0.75$.

\subsection{Dynamic variance for the system of two different oscillators}

Below threshold, the spectra of the $X$ and $Y$ quadratures can be calculated from Eq. \ref{sistema}, translated into the Fourier space. Without parametric driving (with $\barg = 0$), the spectrum of $X$ is 
\begin{equation}
S^0_X (\omega)= \frac{S_{F1}}{2} \frac{1}{(m_1 \omega_1)^2} \frac{1}{\omega^2+\left(\frac{\gamma_1}{2}\right)^2}
\label{SX}
\end{equation}
and its variance is
\begin{equation}
(\sigma_X^0)^2 = \int_{-\infty}^{\infty} S_X^0(\omega)\,\frac{\ud \omega}{2 \pi}\,=\,\frac{S_{F1}}{2 m_1^2 \omega_1^2 \gamma_1}
\label{sigmaX}
\end{equation}
For the $Y$ quadrature, the spectrum and the variance are given by Eqs. (\ref{SX}) and (\ref{sigmaX}) by exchanging the subscripts $1 \leftrightarrow 2$. Before analyzing the effect of the parametric drive, we normalize the quadratures to the respective standard deviation calculated at $\barg=0$. We thus define, similarly to what we have done for the spectral densities, $\tX = X/\sigma_X^0$ and $\tY = Y/\sigma_Y^0$. In the strongly quantum limit, $\sigma^0_X = \sqrt{\hbar/\omega_1 m_1}$ and $\sigma^0_Y = \sqrt{\hbar/\omega_2 m_2}\,$ : our normalization procedure corresponds to considering the usual dimensionless variables of the oscillators, normalized to the zero-point fluctuations.

The evolution equation $\dot{\widetilde{\bf{X}}} + \widetilde{\bf{A}} \widetilde{\bf{X}} = \widetilde{\boldsymbol{\xi}}$ is now characterized by the system matrix
\begin{equation}
\widetilde{\bf{A}}=\left(\begin{array}{cc}
\frac{\gamma_1}{2}\, & \, \frac{\gamma_1}{2} \barg \alpha' \\
 \frac{\gamma_2}{2} \barg/\alpha' \, &\, \frac{\gamma_2}{2}
 \end{array} \right)
 \label{eqtA}
\end{equation}
where now the asymmetry parameter is
\begin{equation}
\alpha' = \sqrt{\frac{S_{F2} m_1 \omega_1}{S_{F1} m_2 \omega_2}}
\end{equation}
that becomes $\alpha' \simeq \sqrt{Q_1/Q_2}$ in the classical limit and $\alpha' \simeq \sqrt{\gamma_2/\gamma_1}$ in the strongly quantum limit, and by the noise vector $\widetilde{\boldsymbol{\xi}} = \left(\sqrt{\frac{\gamma_1}{2}} f_1, \sqrt{\frac{\gamma_2}{2}} f_2 \right)$.   

Below threshold, we can still calculate the stationary spectra
\begin{equation}
\begin{array}{rcl}
S_{\tX \tX} &=& \left[ \left( \omega^2+\frac{\gamma_2^2}{4}\right)\gamma_1+\frac{\gamma_1^2}{4}\barg^2 \alpha'^2 \gamma_2\right]/\mathcal{D}  \\
S_{\tY \tY} &=& \left[ \left( \omega^2+\frac{\gamma_1^2}{4}\right)\gamma_2+\frac{\gamma_2^2}{4}\frac{\barg^2}{\alpha'^2} \gamma_1\right]/\mathcal{D}  \\
S_{\tX \tY} + S_{\tY \tX} &=& -\frac{1}{\mathcal{D}} \frac{\gamma_1 \gamma_2}{2} \barg (\gamma_1 \alpha' + \gamma_2/\alpha')  \\
\mathcal{D} &=& \left(\omega^2-\frac{\gamma_1 \gamma_2}{4}\left(1-\barg^2\right)\right)^2+\omega^2\frac{(\gamma_1+\gamma_2)^2}{4}
\end{array}
\end{equation}
The maximally squeezed quadrature and the corresponding variance are derived by first calculating the spectral integrals $\sigma^2_{\tX \tX} = \int_{-\infty}^{\infty} S_{\tX \tX}\,\frac{\ud \omega}{2 \pi}$ (and similar expressions for $\sigma^2_{\tY \tY}$ and $\sigma^2_{\tX \tY}$), then using Eqs. (\ref{thmin0}) and (\ref{Smin0}) where the spectral densities are replaced by the respective covariance matrix elements.

A different strategy for calculating the time evolving variance can be applied even above threshold, starting from the formal solution of the evolution equation, that reads
\begin{equation}
{\bf \tX}(t)=\bphi(t){\bf \tX}(0) + \int_0^t \bphi(t-\tau)\bxi(\tau) \ud\tau  
\end{equation}
where 
\begin{equation}
\bphi(t) = {\bf M}\left(\begin{array}{cc}
e^{-\lambda_+ t} & 0 \\
 0  & e^{-\lambda_- t}
\end{array} \right) {\bf M}^{-1}
\end{equation} 
and ${\bf M}$ is the matrix of the eigenvectors of $\widetilde{{\bf A}}$, whose eigenvalues $\lambda_{\pm}$ are given explicitly in Eq. (\ref{eigen}). The elements of the covariance matrix for the vector ${\bf \tX}$ have the following temporal evolution
\begin{eqnarray}
\sigma^2_{\tX \tX} (t) &= & \bphi_{11}^2 (t) \sigma^2_{\tX \tX} (0)+ \bphi_{12}^2 (t) \sigma^2_{\tY \tY} (0) \nonumber\\
& & + \gamma_1 \int_0^t \bphi^2_{11}(t-s) \ud s + \gamma_2 \int_0^t \bphi^2_{12}(t-s) \ud s    \\
\sigma^2_{\tX \tY} (t) & = & \bphi_{11} (t) \bphi_{21} (t)\sigma^2_{\tX \tX} (0)+ \bphi_{12} (t) \bphi_{22} (t)\sigma^2_{\tY \tY} (0) \nonumber\\
& &+ \gamma_1 \int_0^t \bphi_{11}(t-s) \bphi_{21}(t-s) \ud s + \gamma_2 \int_0^t \bphi_{12}(t-s) \bphi_{22}(t-s) \ud s 
\end{eqnarray}
($\sigma^2_{\tY \tY}$ and $\sigma^2_{\tY \tX}$ are found from the above equations exchanging $\tX \leftrightarrow \tY$ and $1 \leftrightarrow 2$). As in the stationary case, the maximally squeezed quadrature and the corresponding minimal variance are derived using Eqs. (\ref{thmin0}) and (\ref{Smin0}) where the spectral densities are replaced by the respective covariance matrix elements.

\end{document}